\documentclass[amsmath,amssymb,aps,twocolumn,longbibliography,notitlepage]{revtex4-1}
\usepackage{graphicx}
\usepackage{verbatim}
\usepackage{subfigure}
\usepackage{hyperref}

\newcommand{\tj}[9]{ \begin{pmatrix}
  #1 & #2 & #3 \\
  #4 & #5 & #6
 \end{pmatrix}}

\newcommand{\Gj}[6]{ \begin{Bmatrix}
  #1 & #2 & #3 \\
  #4 & #5 & #6
 \end{Bmatrix}}

\begin{document}

\title{High-resolution optical spectroscopy with a buffer-gas-cooled beam of BaH molecules}



\author{G. Z. Iwata, R. L. McNally, and T. Zelevinsky}
\email{tanya.zelevinsky@columbia.edu}
\affiliation{Department of Physics, Columbia University, 538 West 120th Street, New York, NY 10027-5255, USA}

\begin{abstract}
Barium monohydride (BaH) is an attractive candidate for extending laser cooling and trapping techniques to diatomic hydrides.  The apparatus and high-resolution optical spectroscopy presented here demonstrate progress toward this goal.  A cryogenic buffer-gas-cooled molecular beam of BaH was constructed and characterized.  Pulsed laser ablation into cryogenic helium buffer gas delivers $\sim1\times10^{10}$ molecules/sr/pulse in the X$^2\Sigma^+$ ($v''=0,N''=1$) state of primary interest.  More than $1\times10^7$ of these molecules per pulse enter the downstream science region with forward velocities below 100 m/s and transverse temperature of 0.1 K.  This molecular beam enabled high-resolution optical spectra of BaH in quantum states relevant to laser slowing and cooling.  The reported measurements include hyperfine structure and magnetic $g$ factors in the X$^2\Sigma^+$, B$^2\Sigma^+$, and A$^2\Pi_{1/2}$ states.
\end{abstract}

\maketitle

\section{Introduction}

Experiments with cold and ultracold molecules are at the forefront of table-top fundamental physics \cite{DeMillePT15_FundamentalPhysicsWithDiatomicMolecules} and studies of quantum gases \cite{YeMosesNPhys17_PolarMoleculeFrontiers}.  Diatomic molecules offer many features that are absent in cold atomic gases, such as dipole moments and rich dynamics of vibration and rotation.  These properties make them appealing for a wide range of experiments including measurements of fundamental symmetries \cite{ACMEScience14_ElectronEDM}, investigations of anisotropic quantum gases \cite{ReyHazzardPRL14_ManyBodyDynamicsPolarMolecules}, and demonstrating the quantum mechanical nature of basic chemical processes \cite{ZelevinskyMcDonaldNature16_Sr2PD}.

To enable control of diatomic molecules at the quantum state level, they must be trapped and cooled to ultracold temperatures, generally well below a millikelvin.  In atomic physics experiments, the workhorse of ultracold science is the magneto-optical trap (MOT), which confines and simultaneously cools the particles with magnetic field gradients and optical cycling using laser light.  While MOTs work best with two-level systems, there exists a set of diatomic and even larger molecules where one can isolate a quasi-two-level system despite the rovibrational level complexity, and carry out laser cooling and trapping \cite{DiRosaEPJD04_LaserCoolingMolecules,YeStuhlPRL08_PolarMoleculeMOT}.  These ideas have enjoyed considerable success in the past several years, resulting in molecular MOTs of strontium monofluoride (SrF) and calcium monofluoride (CaF) \cite{DeMilleSteineckerCPC16_ImprovedSrF_RF_MOT,TarbuttTruppeArXiv17_CaFBelowDopplerLimit,DoyleAndereggArXiv17_CaF_RF_MOT} and a two-dimensional MOT of yttrium monoxide (YO) \cite{YeHummonPRL13_YO_MOT}.  Other molecular species under active investigation include strontium monohydroxide (SrOH) \cite{DoyleKozyryevPRL17_SrOHSisyphusCooling} and barium monofluoride (BaF) \cite{YanBuPRA17_BaFBufferGasSpectrosc}.

A number of diatomic hydrides also offer promising laser-cooling schemes \cite{DiRosaEPJD04_LaserCoolingMolecules,LanePRA15_HFromBaH}, but remain underexplored despite their amenability to theoretical treatment.  A cold and slow beam of CaH has been reported \cite{DoyleLuPCCP11_SlowCaHBeam} and cold beams of LiH have been investigated \cite{TarbuttTokunagaJCP07_LiHSupersonicBeam}, both based on laser ablation of solid targets, but suffered relatively low molecule yields compared to some of the diatomic fluorides and oxides \cite{DoyleHutzlerCR12_BufferGasBeams}.  In this work, we demonstrate a bright and slow beam of barium monohydride (BaH) molecules with good stability and sufficiently high flux for optical slowing and cooling.  BaH molecules present the challenges of insufficient spectroscopic information and the usual difficulty of hydride ablation, as well as a large mass, relatively long wavelengths of the optical cycling transitions, and naturally low photon scattering rates, all three of which suppress laser-cooling efficiency.  On the other hand, BaH opens the door to ultracold diatomic hydrides, presents a low Doppler cooling limit $<40$ $\mu$K, and offers an opportunity to test new molecular MOT schemes due to its unusual magnetic moments.  Further, it has a very large mass ratio of the constituent atoms which, from kinetic considerations, could yield ultracold dilute hydrogen samples upon dissociation \cite{LanePRA15_HFromBaH}.

The optical transitions that can potentially support laser cooling are X$^2\Sigma^+$ $\leftrightarrow$ A$^2\Pi_{1/2}$ at 1061 nm and X$^2\Sigma^+$ $\leftrightarrow$ B$^2\Sigma^+$ at 905 nm.  The former transition is analogous to the ones used in most other molecular laser-cooling experiments \cite{DeMilleShumanNature10_SrFLaserCooling,TarbuttTruppeArXiv17_CaFBelowDopplerLimit,YeHummonPRL13_YO_MOT}, while the latter transition could allow an alternative optical cycling scheme \cite{TarbuttTruppeArXiv17_CaFBelowDopplerLimit}.  Typically, the X$^2\Sigma^+$ $\leftrightarrow$ B$^2\Sigma^+$ transition involves a significant hyperfine splitting in the excited state which makes it difficult to ensure appropriate cooling laser detunings from all hyperfine levels.  The X$^2\Sigma^+$ $\leftrightarrow$ A$^2\Pi_{1/2}$ transition is thus more commonly used, but this excited state tends to have a very small magnetic moment, leading to weak magneto-optical trapping forces \cite{TarbuttNJP15_ComplexLevelMOT}.  In BaH, both of these challenges are mitigated.  As we show here, the magnetic moment of the A$^2\Pi_{1/2}$ state is of the same magnitude as that of the ground state, permitting strong trapping forces.  In addition, we have measured a relatively large hyperfine splitting of the B$^2\Sigma^+$ state, and this feature, combined with a narrow 1.6 MHz natural linewidth of this transition \cite{BergPS97_BaHLifetime}, is favorable for efficient magneto-optical trapping on the X$^2\Sigma^+$ $\leftrightarrow$ B$^2\Sigma^+$ transition as well.

In Sec. \ref{sec:Apparatus} of this work, we discuss the details of the cryogenic molecular beam source.  In Sec. \ref{sec:Beam}, we describe the properties of the resulting molecular beam.  Section \ref{sec:Hyperfine} presents hyperfine-structure-resolved optical spectroscopy of the cryogenic beam, while Sec. \ref{sec:gFactors} shows spectroscopy with applied magnetic fields and measurements of relevant magnetic $g$ factors.  In Sec. \ref{sec:Outlook}, we summarize the results and present conclusions.

\section{Experimental apparatus}
\label{sec:Apparatus}

\begin{figure}[h]
\includegraphics[width=8.1cm]{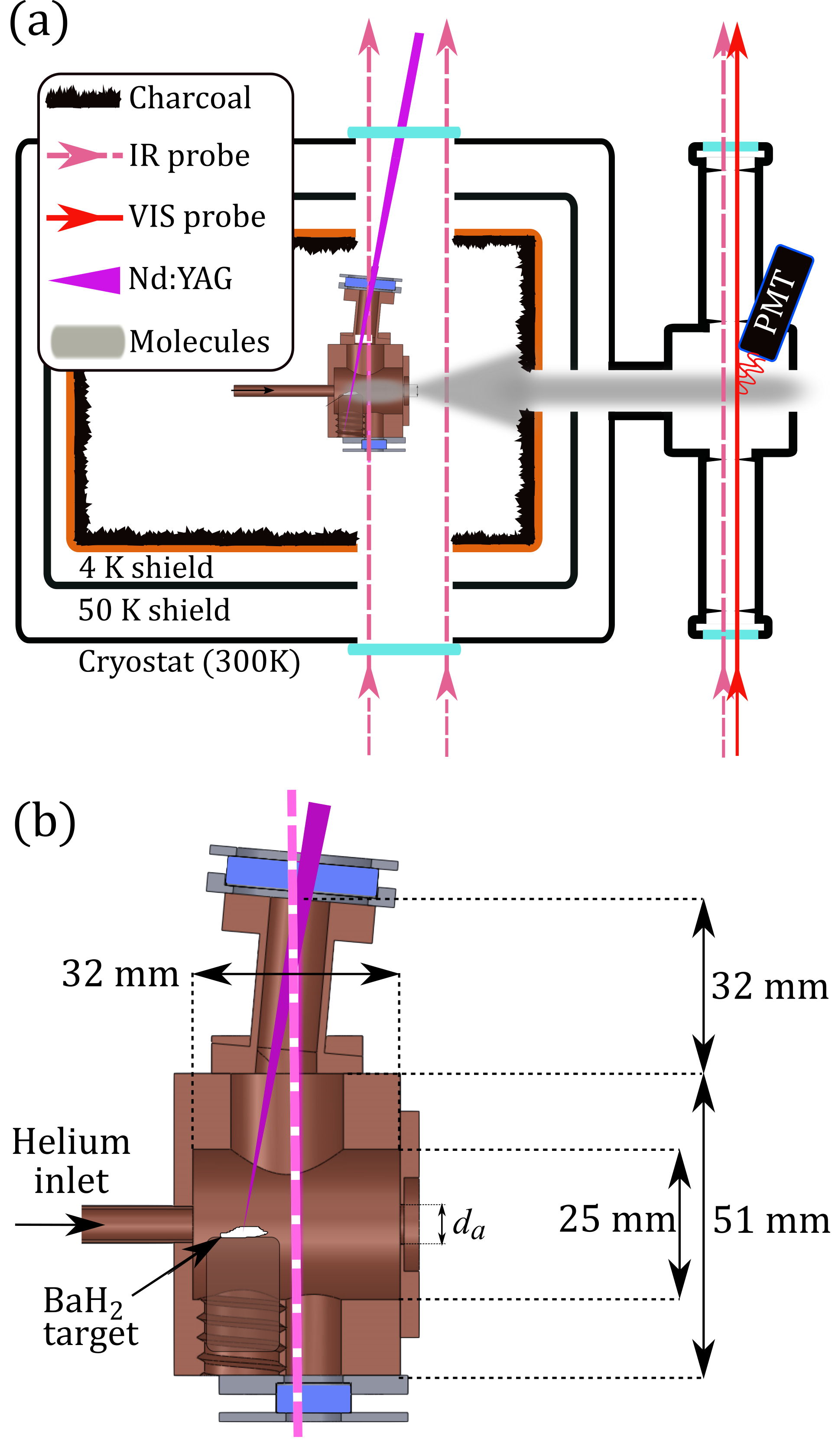}
\caption{(a) Diagram of the buffer-gas-cooled beam source of BaH.  The diatomic molecules are created through pulsed laser ablation of a BaH$_2$ rock inside a cryogenic copper cell filled with flowing He gas.  As the molecules are swept out of the cell, they can be optically probed at various locations along the beam:  in the cell and just after the cell via absorption, and in the downstream science region which is furnished with a PMT for fluorescence detection using visible light.  Coconut charcoal coating on the inner shield as well as on additional copper fins (not shown) acts as a fast cryopump for excess He.  The inner copper shield is nominally at 4 K and is surrounded by a 50 K aluminum shield.  The entire assembly is enclosed in a vacuum-tight aluminum chamber with large optical windows on both sides.  (b) Details and dimensions of the cryogenic buffer-gas cell.  The beam aperture $d_a$ was varied from 3 to 7 mm.}
\label{fig:app}
\centering
\end{figure}

The cryogenic molecular beam source of BaH is shown in Fig. \ref{fig:app}(a).  BaH molecules are produced via laser ablation of a solid BaH$_2$ target inside a copper cell held at $\sim6$ K.  The cell is at the center of a cryostat cooled by a pulse tube refrigerator (PTR).  Helium buffer gas flows into the cell and past the target at $\sim10$ sccm via an inlet pipe at the back of the cell.  The ablated molecules thermalize with the He before being swept out of a $\sim5$-mm-diameter beam aperture.  In order to maintain vacuum in the cryogenic source region ($\sim10^{-6}$ torr while He is flowing), the interior of the 4 K shield is coated with activated coconut charcoal (Calgon Carbon OLC $12\times30$) which acts as a fast cryopump for He.  The charcoal is attached by liberally coating the copper shields with thermal epoxy (Loctite Stycast 2850FT).  The charcoal-coated panels are regenerated by warming up the source region to $\sim80$ K every 10 hours of active operation.  In addition, the panels are weekly cycled to room temperature and occasionally baked to $>100$ $^{\circ}$C.  The source region takes $<14$ hours to cool down to the minimum cell temperature or to fully warm up to room temperature.

The buffer-gas cell, shown in Fig. \ref{fig:app}(b), consists of a cylindrical cavity (1.25 in. long, 1 in. diameter) cut out of a solid copper block.  Smaller cutouts with windows provide optical access for ablation as well as for absorption spectroscopy to monitor the molecule production process.  A woven copper cloth ($100\times100$ mesh size) glued to the helium inlet separates the inlet from the interior of the cell.  To prevent dust buildup on the ablation window, it is extruded by an extended narrow section.  The target holder can screw in and out of the cell wall to facilitate quick and minimally invasive ablation target changes.  Pulsed ablation of a typical target \cite{ZelevinskyTaralloPRA16_BaH} yields approximately 1000 shots, after which the signal begins to deteriorate but can be recovered by manually moving to a different spot on the same target.  The solid targets are produced by cleaving commercial BaH$_2$ rocks 1-3 mm across (Materion DP-532-12-1) to expose a smooth surface, and gluing 4-6 of them to the holder with enough superglue (Loctite 414) to wick up the sides of the rocks.  The rocks are chosen for their flatness and no significant difference in ablation yield was observed after annealing the rocks.  The Nd:YAG ablation laser (BigSky Ultra) emits 8 ns, 50 mJ pulses at 1064 nm with a 1-2 Hz rate and is focused to a 0.5-mm-diameter spot on the target.

We have observed that the rotational temperature of BaH in the buffer-gas cell thermalizes to $\sim9$ K with a time constant of 0.3 ms for typical buffer-gas flow rates and a 7-mm-diameter beam aperture.  The extraction of BaH from the cell occurs between 0.5 and 2 ms after ablation, ensuring good rotational thermalization of the molecules.  This equilibrated rotational temperature is optimal for maximizing population in the X$^2\Sigma^+$ ($v''=0,N''=1$) state of interest for laser slowing and cooling.  Furthermore, the relatively quick extraction times can be beneficial for laser slowing \cite{TarbuttTruppeNJP17_CaFChirpedLaserSlowing}.

Directly outside the buffer-gas cell, molecular beam extraction can be quantified via absorption spectroscopy.  Further downstream,
geometric apertures imposed on the molecular beam consist of a charcoal-coated plate with a 1-cm-diameter hole located 11 cm from the cell exit, and a 3.8-cm-diameter vacuum flange at the output port of the cryostat 53 cm from the cell.  Here the molecules enter a fluorescence-detection region made by a six-way 2.75 in. ConFlat vacuum cube with a black coating (Acktar Metal Velvet) and blackened copper baffles to reduce light scattering \cite{DeMilleNorrgardRSI16_InVacuumSurfaceBlackening}.  This region is pumped by a 700 l/s turbomolecular pump (550 l/s for He), and the vacuum here is currently limited to $\sim10^{-6}$ torr.

The optical cycling transitions for BaH lie in the near-infrared, which is an inconvenient wavelength range for fluorescence detection with photomultiplier tubes (PMTs).  As an alternative, molecules are detected via a higher-energy transition, X$^2\Sigma^+$ $\leftarrow$ E$^2\Pi_{1/2}$, with a 684 nm wavelength.  The 8.6 GHz spin-rotation splitting of the $N''=1$ ground-state rotational level \cite{ZelevinskyTaralloPRA16_BaH} cannot be easily addressed by a single laser, and without repumping and remixing of dark magnetic sublevels, the expected detection rate is about one photon per molecule, even for quasicycling transitions.  However, this is sufficient for precisely characterizing molecular beam parameters and spectroscopic properties.  For measurements of hyperfine energy splittings and magnetic $g$ factors, probe lasers on the X$^2\Sigma^+$ $\rightarrow$ A$^2\Pi_{1/2}$ or X$^2\Sigma^+$ $\rightarrow$ B$^2\Sigma^+$ transitions interact with the molecular beam a few millimeters upstream from the 684-nm detection laser, allowing for resonant depletion or enhancement of the populations in the detected ground states.

\section{Molecular beam}
\label{sec:Beam}

In the context of creating samples of trapped ultracold molecules, the most important molecular beam properties are the flux, forward velocity distribution, and transverse temperature.

The molecular flux downstream from the source is detected via fluorescence with a PMT, as shown in Fig. \ref{fig:app}(a).  For characterizing the molecular beam, we use a single-frequency laser resonantly driving the $v''=0,N''=1,J''=1/2$ spin-rotation level of the electronic ground state X$^2\Sigma^+$ to the $v'=0,N'=0,J'=1/2$ level of the E$^2\Pi_{1/2}$ excited state. Assuming that the molecules are equally distributed in each hyperfine magnetic sublevel, we detect one-third of the total molecules present in the $N''=1$ rotational state.  In addition, the probe laser intercepts slightly less than one-tenth of the molecules in the beam.  The detection efficiency of the system is the product of the 2\% PMT quantum efficiency at 684 nm and the 2.5\% geometric collection efficiency of the detection optics.  All of these factors combined with the signal size of $\sim500$ PMT counts in the detection region per ablation pulse yield approximately $4\times10^7$ molecules in the X$^2\Sigma^+$ ($v''=0,N''=1$) state per pulse.  Some of our ablation targets yield up to three times as many molecules.  Our molecule detection method is sensitive to the transverse Doppler velocity distribution of the beam, and we may be interacting with as few as $10\%$ of the velocity groups during detection.  Since this factor is not included, our molecule count is likely to be significantly underestimated.

\begin{figure}[h]
\includegraphics[trim=0in 0in 0in 0in,clip,width=8.6cm]{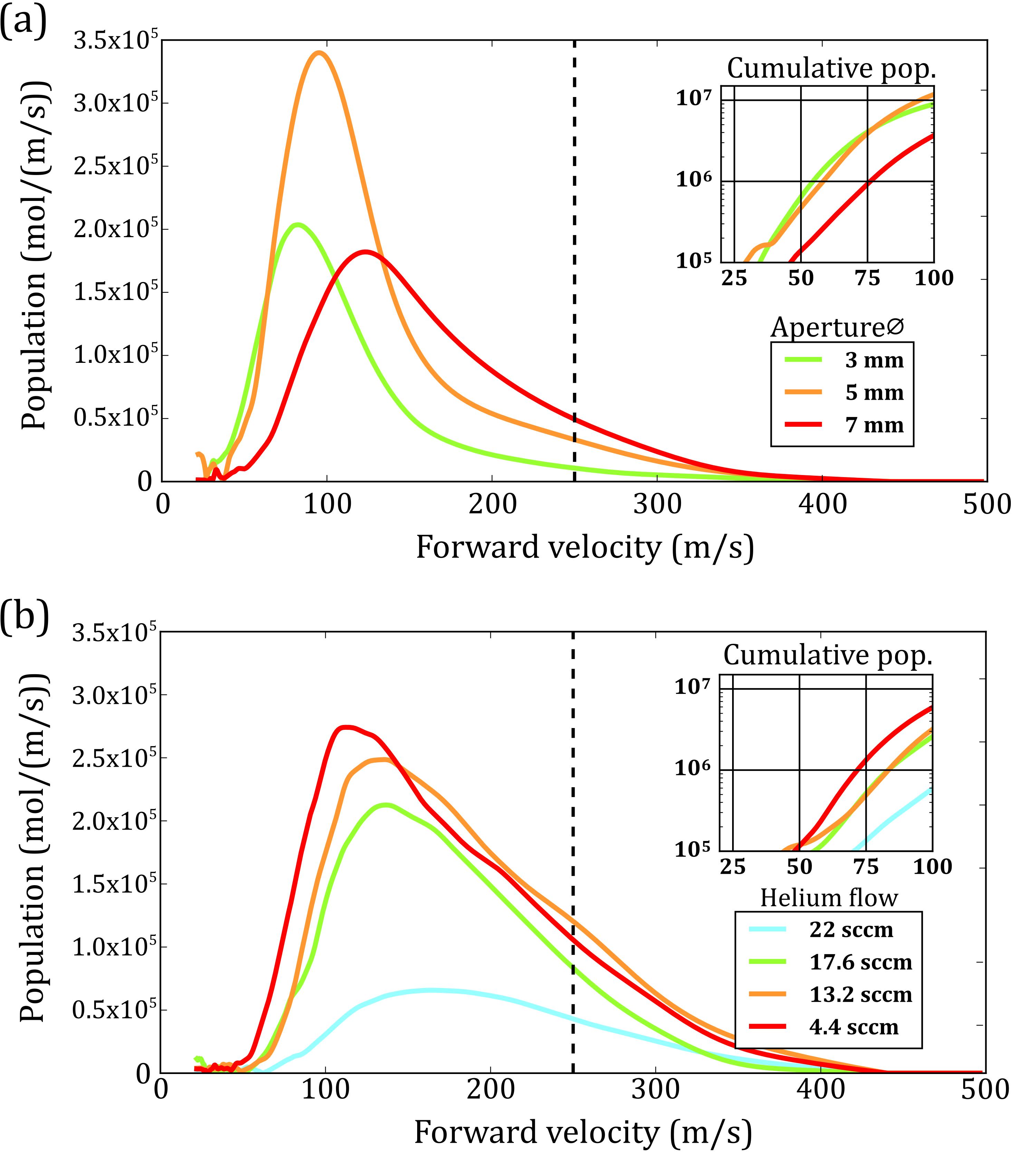}
\caption{Measured forward-velocity distributions of BaH molecules for a range of beam-aperture diameters and buffer-gas flow rates.  Velocity distributions above $\sim250$ m/s (dashed lines) are slightly less reliable due to a mild sensitivity on specific data cuts made to reject fluorescence noise from ablation light.  The small-scale structure at low velocities is an artifact of deconvolution. (a) Velocity distributions for three beam-aperture diameters.  The inset shows a conservatively estimated cumulative population of molecules as a function of forward velocity under 100 m/s.  The 5 mm aperture is optimal, although a slight enhancement of very slow molecules at $\sim50$ m/s is observed for the 3 mm aperture.  For this data set, He flow rates of 11$\pm$2 sccm were used.  (b) Velocity distributions for four He flow rates and a 7 mm cell aperture, with the inset as in part (a).  There is an enhancement of slow molecules at lower flow rates.}
\label{fig:vel_dist}
\centering
\end{figure}
To determine the forward-velocity distribution of the molecules, we make time-resolved molecular density measurements 2 cm away from the cell exit via absorption spectroscopy and, simultaneously, downstream in the fluorescence-detection region, using the X$^2\Sigma^+$ ($v''=0,N''=1$) ground state.  These two measurements of beam density as a function of time are smoothed and numerically deconvolved to yield the forward-velocity distribution in a process analogous to spatial time-of-flight analysis in quantum gas experiments.  The data smoothing is critical for obtaining faithful velocity distributions since numerical deconvolution can amplify high-frequency noise in the data and give nonphysical results, and does not affect the integrated cumulative molecular populations.  The results of this analysis are given in Fig. \ref{fig:vel_dist}(a) for several beam-aperture diameters from 3 to 7 mm and in Fig. \ref{fig:vel_dist}(b) for various He flow rates from 4.4 to 22 sccm.  The insets show cumulative molecule numbers detected downstream with slow forward velocities below 100 m/s.  The 5 mm cell aperture is optimal, although the 3 mm design results in slightly improved numbers of very slow molecules near 50 m/s.  Furthermore, we observe that the number of slow molecules with forward velocities below 100 m/s increases substantially as the He flow is reduced from 22 to 4.4 sccm.  At even lower flow rates, the molecular flux begins to degrade.
The velocity distributions above $\sim250$ m/s are slightly less reliable since this signal comes from faster molecules arriving in the detection region very shortly after ablation, while there is a minimum waiting time after ablation to begin detecting in order to avoid the stray fluorescence.  The relatively broad and asymmetric nature of the velocity distributions for the 7 mm cell aperture is the result of incomplete thermalization of ablated BaH with cryogenic He.  In addition, for the cell with a 5 mm beam aperture, we varied the length of this circular aperture from $<0.5$ mm to 3 mm.  We found that while the overall molecular flux was larger for the longer aperture, the slow molecule flux was several times greater for the short aperture, which was used for all data in Fig. \ref{fig:vel_dist}.  Thus, by optimizing the cell geometry and the buffer-gas flow rate, we detect a conservatively calculated number of $>10^7$ molecules in the fluorescence-detection region with forward velocities $<100$ m/s.

In the fluorescence-detection region, we can limit the transverse temperature of the molecules to 0.1 K by comparing the expected natural linewidth of the B$^2\Sigma^+$ state \cite{BergPS97_BaHLifetime} to the measured spectra.  This cold transverse temperature is consistent with geometric constraints on the beam and allows us to characterize the relevant properties of BaH at a higher optical resolution than was previously possible, enabling direct measurements of hyperfine structure and molecular $g$ factors.

\section{Hyperfine-structure measurements}
\label{sec:Hyperfine}

To explore the possibility of optical radiation pressure experiments with BaH, it is necessary to understand the ground- and excited-state hyperfine structure.  While BaH has been studied intermittently for over 100 years, it is less well explored than other molecular candidates for laser cooling.  The cold molecular beam described here is well suited for optical spectroscopy that reveals hyperfine structure of the ground and excited states.

The transverse Doppler width of the molecular beam in the fluorescence-detection region, in combination with our signal-to-noise ratio, allows us to resolve hyperfine energy splittings at the $\sim4$ MHz level.  It is challenging to directly detect fluorescence from the A$^2\Pi_{1/2}$ or B$^2\Sigma^+$ excited states due to the PMT wavelength limitations.  However, these states can be probed by optically pumping molecules between the two ground-state spin-rotation levels ($J''=3/2$ and $J''=1/2$), while detecting population in one of them via the X$^2\Sigma^+$ $\leftarrow$ E$^2\Pi_{1/2}$ transition.  This is possible due to the highly favorable vibrational branching ratios of the A$^2\Pi_{1/2}$ and B$^2\Sigma^+$ excited states into the ground state \cite{ZelevinskyTaralloPRA16_BaH,LanePRA15_HFromBaH}.

\begin{figure}[h]
\includegraphics[width=8.6cm]{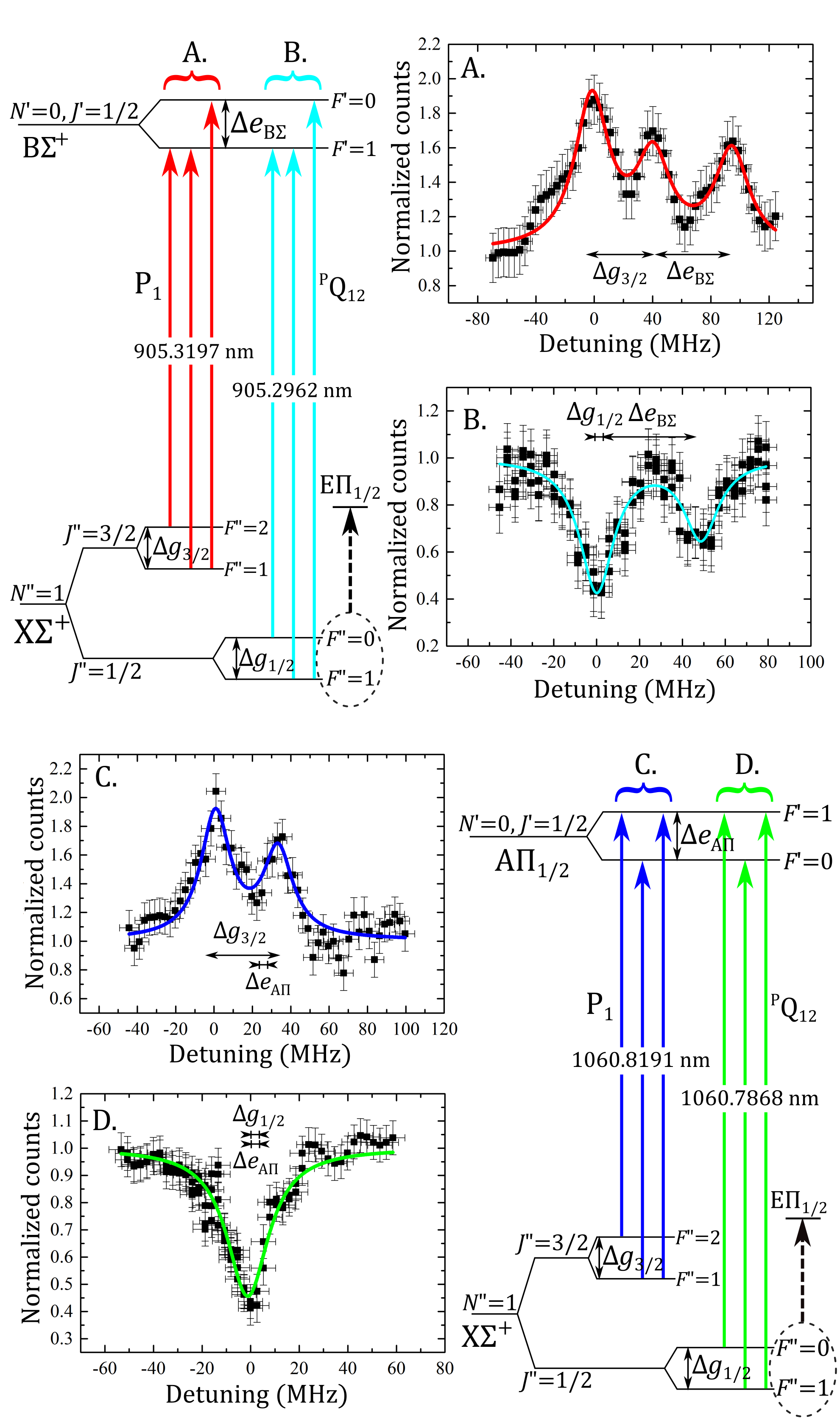}
\caption{Hyperfine-structure-resolved energy levels and measurements for BaH electronic states relevant to laser cooling.  Four types of spectra are collected (labeled A, B, C, and D).  In all cases, detection is made by monitoring the fluorescence of molecules excited from the X$^2\Sigma^+ (N''=1, J''=1/2)$ ground state to the E$^2\Pi_{1/2}$ excited state, indicated by the dashed arrows.  Transitions labeled A and C show population enhancement due to pumping from the X$^2\Sigma^+ (N''=1, J''=3/2)$ state via B$^2\Sigma^+$ or A$^2\Pi_{1/2}$.  Transitions labeled B and D show population depletion due to pumping out of the X$^2\Sigma^+ (N''=1, J''=1/2)$ state.  The studied ground-state and excited-state hyperfine intervals are marked as $\Delta g$ or $\Delta e$ in the diagrams and in the spectra.}
\label{fig:hfine}
\centering
\end{figure}
\begingroup
\begin{table}[h]
\caption{\label{table:hfine} Measured hyperfine intervals for the electronic states of BaH that are relevant to laser cooling, and their estimated uncertainties.  Negative values denote ``flipped" hyperfine structure.}
\begin{ruledtabular}
\begin{tabular}{l c c}
State & Measured hyperfine spacing in MHz\\
\hline
X$^2\Sigma^+(J''=1/2)$ & $-$0(4) \\
X$^2\Sigma^+(J''=3/2)$ & 39(4) \\
\hline
B$^2\Sigma^+(J'=1/2)$ & $-$52(5) \\
\hline
A$^2\Pi_{1/2}(J'=1/2)$ & 0(4) \\
\end{tabular}
\end{ruledtabular}
\end{table}
\endgroup
Hyperfine-structure measurements are outlined in Fig. \ref{fig:hfine} and the results are reported in Table \ref{table:hfine}.  Four types of optical spectra were collected to fix the values of the four unknown hyperfine intervals.  These are labeled A, B, C, and D in Fig. \ref{fig:hfine}.  In each of these experiments, the X$^2\Sigma^+ (J''=1/2)$ population is monitored by sending a 683.7268 nm laser beam through the detection region and recording fluorescence from the spontaneous decay of the E$^2\Pi_{1/2}$ state.  Several beam waists upstream, a probe laser drives transitions from one of the ground electronic spin-rotation levels to the B$^2\Sigma^+$ or A$^2\Pi_{1/2}$ excited state.  The A and C transitions (905.3197 and 1060.8191 nm) enhance the detected fluorescence as the molecules get pumped from the $J''=3/2$ ground-state level to $J''=1/2$, while the B and D transitions (905.2962 and 1060.7868 nm) decrease the fluorescence as the molecules are pumped out of the $J''=1/2$ level.  Each data point in the spectra of Fig. \ref{fig:hfine} is an average of five ablation shots.
The peaks and their spacings in the spectra can be identified by assuming an ordering for the excited-state hyperfine structure and fitting the peak positions. The relative peak amplitudes are given by the hyperfine level degeneracies.  A correctly chosen ordering yields peak spacings and heights that are consistent for all four data sets in Fig. \ref{fig:hfine}.  The shown fits to the fluorescence spectra are constrained only by the expected peak height ratios.

The hyperfine-structure data can be analyzed to extract molecular hyperfine constants which have been previously measured for ground-state BaH in cryogenic solid argon \cite{WeltnerKnightJCP71_AlkalineEarthHydridesHyperfine}.  Excluding negligible terms, the hyperfine structure of X$^2\Sigma^+$ is described by the Hamiltonian
\begin{equation}
H_{\mathrm{hf}} = b_F \mathbf{S}\cdot\mathbf{I} + c I_z S_z
\label{eq:hfint}
\end{equation}
where $b_F$ is the Fermi contact interaction constant, $c$ is the dipolar coupling constant, and $\mathbf{S}$ and $\mathbf{I}$ are the electronic and nuclear spin angular momenta.
Each term in Eq. (\ref{eq:hfint}) can be evaluated as in Sec. 9.5 of Ref. \cite{BrownCarrington}.
The resulting hyperfine interaction matrix elements for $^2\Sigma^+$ states are
\begin{equation}
H_{\mathrm{hf}} =
 \begin{pmatrix}
 \frac{b_F}{4} + \frac{c}{20} & 0 & 0 & 0 \\
 0 & -\frac{b_F5}{12} -\frac{c}{12} & \frac{b_F\sqrt{2}}{3} + \frac{c\sqrt{2}}{6} & 0 \\
 0  & \frac{b_F\sqrt{2}}{3} + \frac{c\sqrt{2}}{6} & -\frac{b_F}{12} +\frac{c}{12} & 0  \\
 0 & 0 & 0 & \frac{b_F}{4} - \frac{c}{4}
 \end{pmatrix}.
 \nonumber
\end{equation}
Our hyperfine-structure measurements yield $b_F = 50(7)$ MHz and $c = 39(8)$ MHz.  The value for $b_F$ is consistent with previous measurements of 47(2) MHz \cite{WeltnerKnightJCP71_AlkalineEarthHydridesHyperfine}.  The value for $c$, while lacking previous reliable measurements, is consistent with those for other alkaline-earth-metal monohydrides \cite{WeltnerKnightJCP71_AlkalineEarthHydridesHyperfine}.

The hyperfine-structure results in Table \ref{table:hfine} can guide experiments on radiation-pressure slowing and cooling of BaH.  For the lower spin-rotation level of the ground state ($J''=1/2$), the hyperfine structure is unresolved, while for the higher level ($J''=3/2$), the splitting is 39(4) MHz and can be easily covered by sidebands imprinted on the laser light with standard electro-optical techniques.  The hyperfine structure is also small, or of the order of the natural linewidth, in the A$^2\Pi_{1/2}$ excited state as is the case for other diatomic molecules that have been investigated as laser-cooling candidates.  This feature allows all excited-state sublevels to participate in optical cycling, thus maximizing radiation pressure forces.  The 52(5) MHz hyperfine interval in the B$^2\Sigma^+$ excited state is $\sim30$ times larger than the natural linewidth and is of a similar magnitude to that of the ground-state $J''=3/2$ level, such that the combination of the two can be managed with electro-optical techniques.

\section{Magnetic $g$-factor measurements}
\label{sec:gFactors}

Magnetic $g$ factors are crucial for understanding magneto-optical trapping forces on molecules.  In particular, the trapping forces depend strongly on the ratios of the $g$ factors in the ground and excited states \cite{TarbuttNJP15_ComplexLevelMOT}.  Here we report predictions and measurements of the relevant magnetic $g$ factors in BaH.

The predictions can be made by diagonalizing the Zeeman interaction Hamiltonian for each molecular state.  For $^2\Sigma^+$ states, the electronic and nuclear Zeeman Hamiltonians are expressed in Eqs. (8.183) and (8.185) of Ref. \cite{BrownCarrington}, and involve 12 hyperfine-structure magnetic sublevels $m_F$ for the X$^2\Sigma^+$ state and 4 sublevels for the B$^2\Sigma^+$ state of BaH.

The Zeeman shifts for $^2\Pi$ states are strongly influenced by the parity-dependent contributions $g_{l}'\approx\frac{p}{2B}$ and $g_{r}^{e'}\approx\frac{q}{B}$, where $p$ and $q$ are the $\Lambda$-doubling constants and $B$ is the rotational constant.  These constants have been measured in BaH for the A$^2\Pi_{1/2}$ excited state \cite{GuntschKoppFysik66_BaHRotationalAXBand} and for the detection state E$^2\Pi_{1/2}$ \cite{VergesFabreJPB87_BaHDeltaState}.  The matrix elements of the applicable Zeeman Hamiltonian are expressed in Eq. (9.71) of Ref. \cite{BrownCarrington}, and the dominant parity dependent term can be described with a single effective $g$ factor as
$g_{\mathrm{eff}}=(g_l'-g_{r}^{e'})/3$.  This contribution alone would result in a $g$-factor value of $-0.27$ for A$^2\Pi_{1/2}$, while a purely semiclassical prediction would yield a value of 0.
Considering additional contributions due to interactions between electronic states results in a predicted $g_{\mathrm{eff}}$ value of -0.44.  In BaH, the five lowest 5d excited states form an interacting complex \cite{BarrowBernardMP89_BaH5dStates} where the strongest mixing is between B$^2\Sigma^+$ and A$^2\Pi_{1/2}$.  This results in an enhancement of the A$^2\Pi_{1/2}$-state $g$ factor and a slight reduction of the B$^2\Sigma^+$-state $g$ factor. However, the five states considered in the prediction here may be insufficient to fully capture all contributions to the A$^2\Pi_{1/2}$-state $g$ factor.

\begin{figure}
\includegraphics[width=8.6cm]{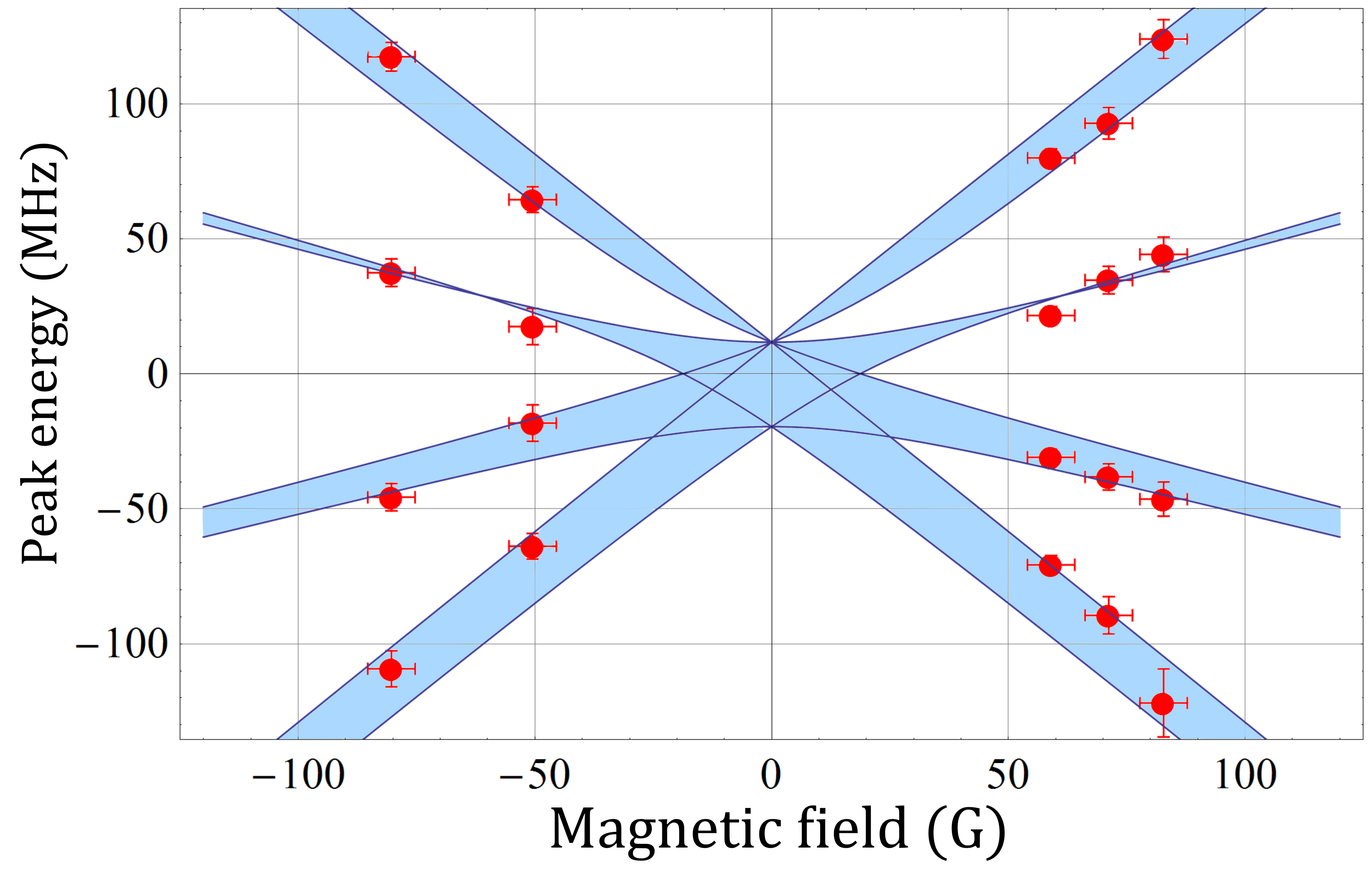}
\caption{Measured Zeeman shifts of the X$^2\Sigma^+(J''=3/2)$ magnetic sublevels, overlayed with a prediction from the Zeeman Hamiltonian.  Solid lines represent energies of the $m_F$ sublevels.  Shaded areas between pairs of sublevels emphasize structure that is spectroscopically unresolved even when selection rules allow both sublevels to couple to the excited state.}
\label{fig:zeeman}
\centering
\end{figure}
\begingroup
\begin{table}
\caption{\label{table:gfactors}Measured and predicted effective $g$ factors in the $m_J$ basis which is most pertinent to the field regimes used in magneto-optical trapping.  The estimated measurement uncertainties are indicated in parentheses.
}
\begin{ruledtabular}
\begin{tabular}{l c c}
State & Measured $g_{\mathrm{eff}}$ & Prediction\\
\hline
$X^2\Sigma^+(N''=1, J''=1/2)$ & $-$0.65(2) & $-$0.70 \\
$X^2\Sigma^+(N''=1, J''=3/2)$ & +0.56(2) & +0.50 \\
\hline
$B^2\Sigma^+(N'=0, J'=1/2)$ & +2.54(6) & +2.86\\
\hline
$A^2\Pi_{1/2}(N'=0, J'=1/2)$ & $-$0.51(2) & $-$0.44\\
\hline
$E^2\Pi_{1/2}(N'=0, J'=1/2)$ & $-$0.04(2) & $-$0.04\\
\end{tabular}
\end{ruledtabular}
\end{table}
\endgroup
While at very low magnetic fields it is natural to use the $m_F$ basis for the Zeeman interaction Hamiltonian, at fields exceeding $\sim10$ G, the Zeeman shifts are best described in the $m_J$ basis.  As a result, the reported factors $g_{\mathrm{eff}}$ describe the measured energy shifts for magnetic field strengths of tens of gauss.  These shifts are $\Delta E=g_{\mathrm{eff}}\mu_B m_J B$, where $B$ is the applied field and $\mu_B$ is the Bohr magneton.
High-resolution Zeeman spectra were collected for both excited states, for both spin-rotation levels of the ground state, and for the E$^2\Pi_{1/2}(N'=0,J'=1/2)$ state used in the detection scheme.  Figure \ref{fig:zeeman} shows Zeeman shift data for the X$^2\Sigma^+(J''=3/2)$ ground-state sublevels, along with the prediction from the Zeeman Hamiltonian.
The experimental results together with the predictions are listed in Table \ref{table:gfactors}.  For all experiments, the magnetic field was applied perpendicularly to the probe laser propagation direction and calibrated \textit{in situ} with a commercial gaussmeter.  Measurements of the ground-state Zeeman shifts were done via the X$^2\Sigma^+$ $\leftarrow$ E$^2\Pi_{1/2}$ transition, where we could separately identify the ground- and excited-state splittings, as well as the relative signs of their $g$ factors, by switching the probe laser polarization between $\pi$ and $\sigma^{\pm}$ transitions.  The Zeeman shifts in the B$^2\Sigma^+$ and A$^2\Pi_{1/2}$ excited states were measured via the population-depletion method as in the hyperfine-structure studies, and polarization was again switched to drive $\pi$ and $\sigma^{\pm}$ transitions in order to determine the relative signs of all the $g$ factors.  To fix the absolute signs, a calibration measurement was made by applying a magnetic field along the laser propagation axis and using circularly polarized light.

The results in Table \ref{table:gfactors} highlight an interesting difference in laser-cooling prospects between BaH and other diatomic molecules currently in use, such as SrF \cite{DeMilleSteineckerCPC16_ImprovedSrF_RF_MOT} and CaF \cite{TarbuttTruppeArXiv17_CaFBelowDopplerLimit,DoyleAndereggArXiv17_CaF_RF_MOT}.  Unlike the fluorides, BaH has a large magnetic moment in both excited states that could be used for optical cycling, B$^2\Sigma^+$ and A$^2\Pi_{1/2}$.  This could allow several approaches to magneto-optical trapping using simpler MOT schemes than for the fluorides \cite{TarbuttNJP15_ComplexLevelMOT}.
The excited-state $g$ factors with magnitudes reported in Table \ref{table:gfactors} place these two optical transitions on an equal footing as possible MOT schemes, with trapping forces potentially $\sim5$ times stronger than those using states with near-zero magnetic moments \cite{TarbuttNJP15_ComplexLevelMOT,TarbuttDevlinNJP16_3DMolecularMOTForces}.

\section{Conclusions}
\label{sec:Outlook}

We have built, characterized, and optimized a molecular beam of BaH that is cooled by cryogenic He buffer gas and delivers packets of molecules to a downstream interaction region at a rate of a few hertz.  More than $1\times10^7$ molecules in each packet are transversely cold (0.1 K) and have forward velocities below 100 m/s.  Improvements to the setup could potentially boost this number further, for example by achieving a higher level of vacuum in the molecular beam region.  Additionally, it is possible that producing sintered BaH$_2$ pellets or other types of targets could allow us to reduce the intensity of the ablation laser, which in turn might lead to more efficient thermalization of BaH with the buffer gas.

The molecule numbers and velocities that were achieved here are a starting point for experiments in which radiation-pressure forces are applied to BaH for laser slowing, cooling, and trapping.  The critical spectroscopic parameters were determined in this work both theoretically and experimentally, since the cold molecular beam allows optical spectroscopy with a resolution of a few megahertz.  Hyperfine structure was measured in the ground electronic state and in both excited states that could be used for optical cycling.  Modest magnetic fields were applied to the molecules to measure all relevant magnetic moments of the ground and excited electronic states.  The resulting effective $g$ factors may enable several promising laser cooling and trapping schemes that could result in relatively large trapping forces.

As the experimental field of ultracold molecules gains momentum, it is important to attempt laser-cooling experiments with different types of molecules, and especially with underexplored species such as diatomic hydrides.  BaH is particularly appealing because of its low Doppler cooling limit, good optical cycling properties \cite{ZelevinskyTaralloPRA16_BaH,LanePRA15_HFromBaH}, unusual excited-state magnetic moments as was shown here, and a large mass ratio between its atomic constituents \cite{LanePRA15_HFromBaH}.  The results of this work are an advance toward these goals.

\begin{acknowledgments}
We thank M. G. Tarallo and L. Abbih for their contributions to this work, and S. Truppe and M. R. Tarbutt for helpful discussions.  We acknowledge partial support by the ONR Grant No. N00014-16-1-2224.  R.L.M. and G.Z.I. acknowledge support by the NSF IGERT Grant No. DGE-1069240.
\end{acknowledgments}


\end{document}